\documentclass[aps,prb,twocolumn]{revtex4}
\usepackage{graphicx}
\usepackage{amsmath}
\usepackage{braket}
\usepackage{hyperref}
\usepackage{subfigure}
\usepackage{color}

\newcommand{\ct}{\cite}
\newcommand{\la}{\lambda}
\newcommand{\bi}{\bibitem}
\newcommand{\be}{\begin{equation}}
\newcommand{\ee}{\end{equation}}
\newcommand{\ba}{\begin{eqnarray}}
\newcommand{\ea}{\end{eqnarray}}

\newcommand{\de}{\delta}

\begin{document}
\title{Exploring chaos in Dicke Model using ground state fidelity and Loschmidt echo}
\author{Utso Bhattacharya}
\affiliation{Department of Physics, Indian Insitute of Technology, 208016, Kanpur}
\author{Sayak Dasgupta}
\affiliation{Department of Physics, Indian Insitute of Technology, 208016, Kanpur}
\author{Amit Dutta}
\affiliation{Department of Physics, Indian Insitute of Technology, 208016, Kanpur}
\begin{abstract}
We study the quantum critical behaviour of the Dicke Hamiltonian, with finite number of atoms and explore the signature of quantum chaos using measures like the ground state fidelity and the Loschmidt echo. We show that both these quantities clearly point to the chaotic nature of the system in the super-radiant phase.
 \end{abstract}
\maketitle
\section{Introduction}
\label{sec_intro}
A classical system is said to be integrable, if the number of independent conserved quantities in the system equals the number of degrees of freedom. The motion of a particle then takes place on a $d$-dimensional tori. Whereas, the absence of symmetries in the system makes the particle trajectory to get delocalised over the whole of the energy surface within a bounded region of the phase space. Such trajectories may have hypersensitivity to initial conditions resulting in chaotic dynamics. Such chaotic dynamics in classical systems are generally characterised by a non-zero Lyapunov exponent which quantifies the exponential divergence of ``nearby" trajectories \cite{strogatz}.

There are two different types of motions in classical Hamiltonian mechanics: regular motion of integrable systems and random motion of non-integrable systems. To understand whether a system is chaotic we look at a cluster of trajectories of a Hamiltonian $H$ originating from nearly same initial conditions in the phase space. In chaotic systems any two trajectories separate exponentially fast with time, while for a regular system the separation varies with a power law involving time ($t$).The linearity of quantum mechanics disallows the phenomenon of chaos in quantum systems \cite{stockmann}. Taking two eigenstates of the Hamiltonian $H$ at slightly separate phase space points; after time $t$, $|\braket{\phi(t)|\psi(t)}|^{2}=|\braket{\phi(0)|\psi(0)}|^{2}$ due to the unitary nature of the time evolution operator 
$U = \exp(-iHt/\hbar)$, hence this direct method of taking overlaps does not work in trying to identify the possibilty of chaos for the corresponding classical Hamiltonian.

The correspondence principle however demands that just like their classical counterparts, exponential sensitivity to initial conditions should also manifest itself somehow in quantum dynamics. That is signatures of chaos can be identified for quantum Hamiltonians which will indicate their classical counterparts to be chaotic \cite{stockmann}. Hence, the Loschmidt Echo (LE) measure giving the overlap of the same wavefunction evolved under two slightly different Hamiltonians was proposed as a way to identify chaos in quantum systems \cite{peres84}. In order to understand the role of the LE in understanding ``quantum" chaos, we study LE and other measures related to and derived from it, on the Dicke Hamiltonian (DH) \cite{dicke}.

The Dicke Model is a system  of ``$N$" interacting 2-level atoms placed in a bosonic cavity (or bath) with a coupling characterised by the parameter $\lambda$. This model is widely studied in quantum optics to understand collective effects. In the limit of an infinite system the model is integrable (solvable) and shows a sharp quantum phase transition. The finite sized system (characterized by a finite number of atoms proportional to $j$) has the transition rounded off, however, it shows a transition from a normal phase (quasi-integrable) to a super-radiant (chaotic) phase; as well understood from the studies of energy-level statistics performed on it \cite{brandes}. We use this finite $j$ case to investigate chaos in this present article.

Emary and Brandes \ct{brandes} used level statistics of the energy eigenvalues of the DH in the finite $j$ case to indicate the presence of chaos. They have used the fact that quantum systems have conserved quantities when their classical counterparts have a high degree of symmetry which leads to degeneracy in the energy spectrum. This enables them to construct a nearest neighbour level-spacing distribution $P(S)$, where $P(S)$ is given by the Poisson Distribution, when such symmetries exist, as $S\rightarrow 0$. Here $S$ is the nearest neighbour level spacing. They call such a quantum spectra, "Quasi-Integrable".The classically chaotic regime is however, devoid of symmetries and hence, the quantum Hamiltonian is non-degenerate and absent of energy level-crossings leading to $P(S)\rightarrow 0$ as $S\rightarrow 0$ giving rise to the Wigner-Dyson distribution ($P_{W}(S)=\pi (S/2)\exp(-\pi S^{2}/4)$).

For finite $j$, appearance of Poisson distribution of $P(S)$ in the normal phase and the Wigner-Dyson distribution of the
same  in the super-radiant phase serves as a good signature for the transition to chaos.
However, the fact that this correspondence between the $P(S)$ and the "chaoticity" of the classical or the quantum Hamiltonian is not general or unique and a good number of exceptions do exist \ct{brandes}.This motivates us to look for other signatures to identify chaos in a more general fashion using two quantum information theoretic measures
namely, the ground state fidelity and the time average of the  Loschmidt Echo. We note that there exists a different approach based on the operator fidelity metric \cite{giorda} which bypasses the need to do a perturbative expansion in
the coupling strength   to generate the eigenstates for the modified Hamiltonian with a shifted parameter value. In our case, on the other hand,  we use a numerical method to obtain the eigenstates in a direct fashion.

In recent years there have been many works those studied  the connection between quantum phase transitions \ct{sachdev99}, 
quantum information \cite{amico08,latorre09} and quantum critical dynamics \cite{dutta10,polkovnikov11}. 
Two important measures which show interesting behaviour close to a quantum critical point  are  
Loschmidt echo \cite{quan06,sharma12,cucchietti07} and the ground state quantum fidelity \cite{zanardi06} (see
review articles \ct{dutta10,gu10}).
Especially the former has been studied extensively in recent years in connection to the dynamics of 
decoherence \cite{damski11,mukherjee12,nag12}, the work statistics \cite{silva08}, equilibration \ct{venuti10} and the   dynamical
phase transition \ct{pollmann10,heyl13}.
Furthermore, the concept of Loschmidt echo (LE) was
proposed in connection  to quantum chaos  \ct{peres84} to describe the hyper-sensitivity of the time evolution of the
system to the perturbations experienced by the surrounding environment; there have been a host of
studies in this direction
\ct{zurek94,karkuszewski02,cucchietti03,jalabert01}. To the best of our knowledge, ours is the first attempt
to understand chaos in the present model through the route of the Loschmidt echo.

The paper is organised in the following manner. In section \ref{sec_model} we discuss the DH briefly, also providing a numerical diagonalisation technique. We then move onto the study of ground-state fidelity in both thermodynamic and finite size limits in section \ref{sec_chaos_through_GS_fid}. We then discuss the LE for DH in section \ref{sec_GroundstateLE} followed by the numerical analysis for the time average of LE in section \ref{sec_TimeavgLE}, before drawing our final conclusions in section \ref{conclusion}.\\ 

\section{The Dicke model: infinite and finite $j$}
\label{sec_model}

We look for the signatures of quantum chaos in the Dicke Hamiltoninian (DH) which describes a single mode bosonic field interacting with an ensemble of $N$  two level atoms \ct{dicke}, given by

\begin{equation}
H = \omega_{0}\sum_{i=1}^{N}s_{z}^{i}+\omega a^{\dagger}a+\sum_{i=1}^{N}\frac{\lambda}{\sqrt{N}}(a^{\dagger}+a)(s_{+}^{(i)}+s_{-}^{(i)})\hspace{0.5cm}[\hbar=1].
\label{eq.DH}
\end{equation}
Here $\omega_{0}$ is the level splitting between the two-level systems. $a^{\dagger}(a)$ is the creation (annihilation) operator for the bosonic field; with $[a^{\dagger},a]=1$. In our case, we consider only a single bosonic mode which interacts with two-level atoms with the interaction strength $\lambda$. The $i$-{th} atom is described by the spin-half operators $\left(s_{k}^{i};k=z,\pm\right)$, obeying the commutation rules $[s_{z},s{\pm}]=\pm s_{\pm}$; and $[s_{+},s_{-}]=2s_{z}$.
The origin of the factor $1/\sqrt{N}$ in the interaction term results from the dipole interaction which is proportional to $1/\sqrt{V}$, where $V$ is the volume of the cavity. Taking into consideration that the density of atoms in the cavity is $\rho =N/V$, we find that the coupling strength is of the form  $\lambda/\sqrt{N}$. The scaling factor $\sqrt{N}$ appearing in the interaction plays an important role for the finite ``size" system.

The DH (Eq.~\ref{eq.DH}) is further simplified by using collective atomic operators,
\begin{equation}
J_{z}\equiv\sum_{i=1}^{N}s_{z}^{(i)};\hspace{4mm}J_{\pm}\equiv\sum_{i=1}^{N}s_{\pm}^{(i)},
\label{eq.ang_mom}
\end{equation}
which obey the usual angular momentum commutation relations.
Here, $j$ is assigned its maximum value $j=N/2$, and this value is constant for a fixed value of $N$. Thus, the $N$ two-level system effectively gets reduced to  a $(2j+1)(=(N+1))$ level system.
The final form of the single-mode DH then looks like,
\begin{equation}
H = \omega_{0}J_{z}+\omega a^{\dagger}a+\frac{\lambda}{\sqrt{2j}}(a^{\dagger}+a)(J_{+}+J_{-})
\label{eq.DH2}
\end{equation} 
The resonance condition, $\omega=\omega_{o}=1$, has been used in the rest of the paper.
The parity operator $(\Pi)$ can be defined here in terms of the total number of excitation quanta $(\hat{N})$ in the system, as
\begin{equation}
\Pi=\exp{\{i\pi\hat{N}\}};\hspace{1cm}\hat{N}=a^{\dagger}a+J_{z}+j,
\label{eq.parity_op}
\end{equation}
Clearly, the operator $\Pi$ can have only two eigenvalues $(\pm 1)$, $N$ being even or odd.
Thus, the DH turns out to be parity conserving as $[H,\Pi]=0$ and, correspondingly the Hilbert-space of the total system is split into two  non-interacting sub-spaces.

The DH shows a QPT in the thermodynamic limit (as $N$ $\rightarrow \infty$) at a critical value of the atom-field coupling strength $(\lambda)$, $\lambda_{c}=\sqrt{\omega\omega_{o}}/2$ where the symmetry associated with the parity operator $(\Pi)$ is broken. The second derivative of the ground state energy per $j$ with respect to $\lambda$ shows a sharp discontinuity at the point $\lambda = \lambda_{c}$ clearly marking occurrence of  a phase transition; this transition
separates the normal phase (for $\la <\la_c$) from the  super-radiant (for $\la > \la_c$). The system in the normal phase is only microscopically excited whereas the super-radiant phase shows macroscopic excitations.

In the finite $j$ limit however, parity symmetry holds  and $\Pi$ continues to be  a good quantum number for all
values of $\la$ and  there is  no discontinuity in the ground state energy per $j$ ($=E_{G}/j$) with respect to $\lambda$ indicating the absence of  a sharp phase transition. However, the finite $j$ results tend to the infinite $j$ (i.e., thermodynamic limit) very rapidly. The system however shows microscopic excitations below $\lambda_{c}$ even for finite $j$ and is macroscopically excited above that value although the crossover from the microscopically excited phase to the macroscopically excited phase is not sharp.  Therefore, one observes that the initially localized wave function for a small but finite $j$ gets delocalized rapidly with a slight increase in $j$. Finally as $j \to \infty$,  the wave function breaks into two lobes (creating degeneracy); the parity symmetry breaks and there is a proper QPT at $\lambda_{c}$ in this limit \ct{brandes}. There is no QPT for a finite $j$ case in the true sense of the term, because the parity symmetry remains intact but there is a crossover at around $\lambda_c$ indicating a transition from a localized (normal) phase to a delocalized (chaotic) phase.

Exact solutions of the DH at finite $j$ do not exist except for $j = 1/2.$
Hence, we make resort to a numerical diagonalisation scheme 
using  the number states of the field $\Ket{n}$ and the Dicke states $\Ket{j,m}$ as our combined basis $\{\Ket{n}\otimes\Ket{j,m}\}$. The  approximation we have to make here  is that  the bosonic Hilbert space is truncated  but always ensuring that it is sufficiently large to be considered as a bath. Finally, diagonalising the DH for finite $j(=5)$, we evaluate and plot the ground state energy $(E_{G}/j)$ and the ground state expectation values of the scaled atomic inversion $<J_{z}>/j$ and the photonic number $<a^{\dagger}a>/j$ as a function of the coupling strength $\lambda$.
 We emphasize that our results match with those produced in the paper by Emary and Brandes\cite{brandes}, in terms of both the phase transition point and behavior at high $\lambda$.

\begin{figure}[h]

\includegraphics[scale=.32]{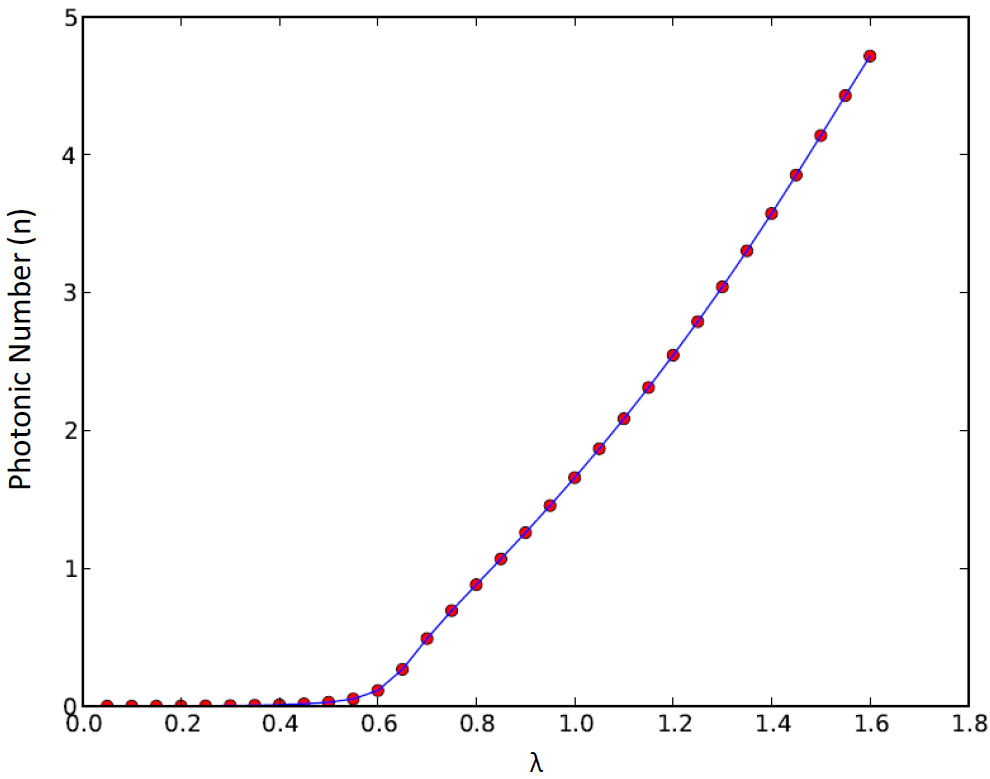}
\begin{picture}(0,0)
\put(-85,28){\includegraphics[height=1.9cm]{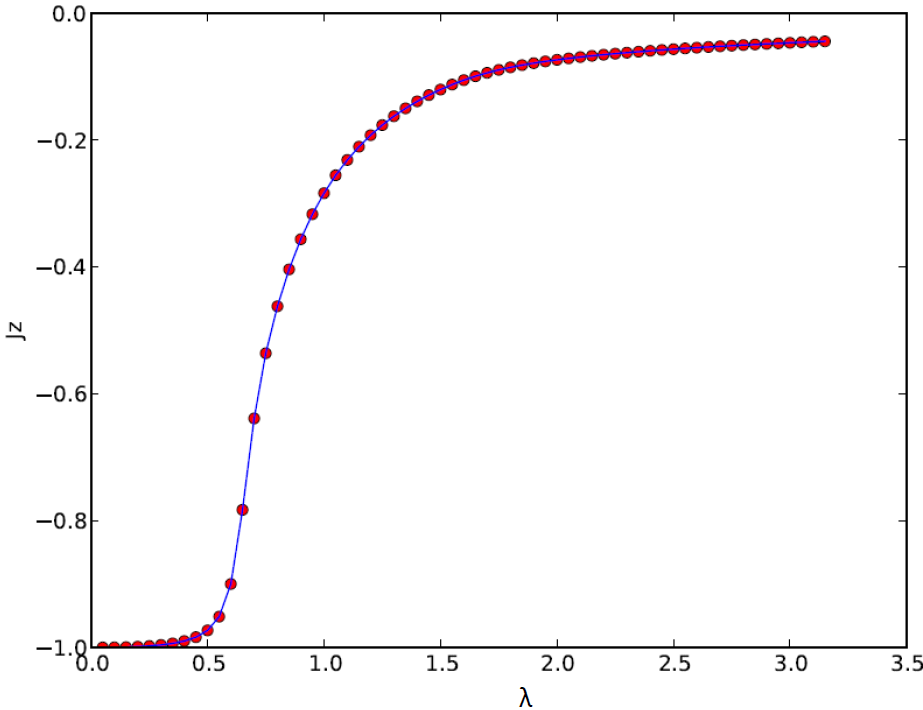}}
\end{picture}
\begin{picture}(0,0)
\put(-218,28){\includegraphics[height=1.9cm]{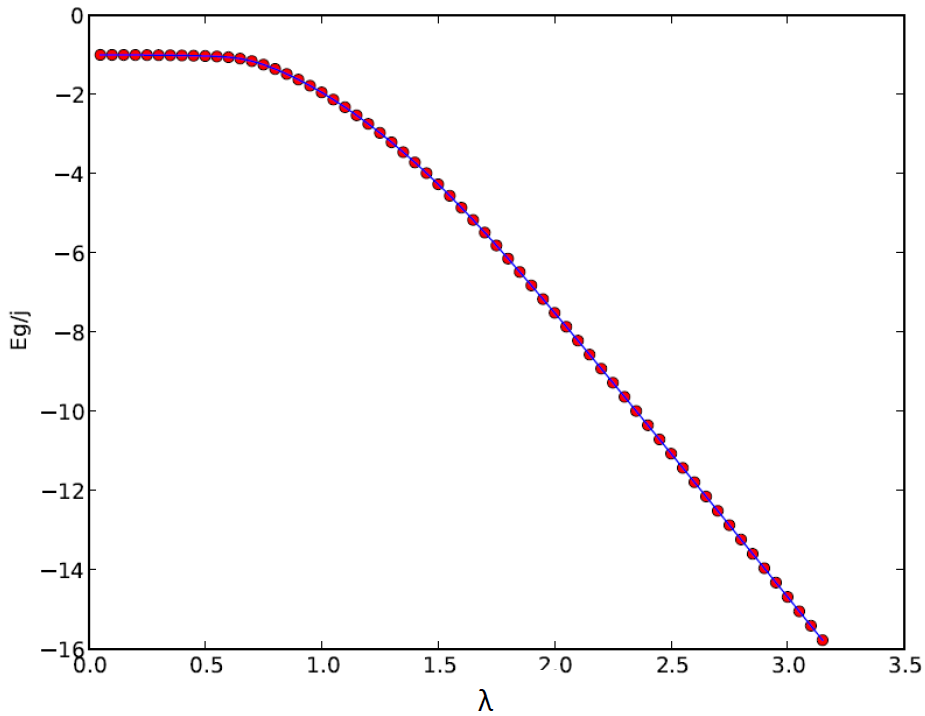}}
\end{picture}

\caption{(colour online) The ground state expectation values of mean photonic number $N$ (background), the ground state energy $E_{g}$(left inset) and the expectation value of the atomic inversion $J_{z}$ (right inset)as a function of $\lambda$ for the numerically diagonalized Hamiltonian.($\lambda_{c} = 0.5$, $\omega = \omega_{o}=1$). The plots produced here match the ones present in ref[\onlinecite{brandes}] and hence establishes the justification of the chosen system parameters $j = 5$ and $n_{c} = 40$ for the remaining simulations.}
\label{fig_expectation_values}
\end{figure}

\section{Identification of Chaos through ground state fidelity}
\label{sec_chaos_through_GS_fid}
The ground state quantum fidelity ($F$), which measures the overlap between many-body ground states at slightly different values of a parameter $\lambda$ of the Hamiltonian usually serves as an important tool for detecting quantum phase transitions. We shall discuss below that it also acts as a  good indicator of transition to quantum chaos. 
Let us define the the ground state fidelity as
\begin{equation}
F = |\braket{\psi(\lambda+\delta)|\psi(\lambda)}|^{2},
\label{eq_fidelity}
\end{equation}
where  $\psi(\la+ \delta)$ and $ \psi(\lambda)$, the ground states of the DH with parameters $\la$ and $\la+\de$, respectively.
We present results for the  ground state fidelity defined in Eq. (\ref{eq_fidelity}) of the DH in both the limits - thermodynamic $(N~ i.e., ~j \rightarrow\infty)$ and finite $j$ ($=5$) in both the phases. Although results obtained  in the thermodynamic
limit were already reported in the reference \onlinecite{zanardi06}, we present them here to highlight the features that emerge in the finite $j$ case, especially in the super-radiant phase.

\subsection{Thermodynamic Limit}
To exactly diagonalise the Hamiltonian in the thermodyanmic limit one resorts to the Holstein-Primakoff representaion of the angular momentum operators, given by :
\begin{eqnarray}
J_{+} &=& b^{\dagger}\sqrt{2j-b^{\dagger}b}\\
 J_{-} &=& \sqrt{2j-b^{\dagger}b}b;\\
J_{z} &=& \left(b^{\dagger}b-j\right)
\label{eq.Holstein_Primakoff}
\end{eqnarray}
where $[b,b^{\dagger}]=1$. With these substitutions we get the DH in the normal phase as:
\begin{eqnarray}
H &=& \omega_{0}\left(b^{\dagger}b-j\right) + \omega a^{\dagger}a \\ \nonumber
&+& \lambda\left(a^{\dagger} + a\right)\left(b^{\dagger}\sqrt{1-\frac{b^{\dagger}b}{2j}}+\sqrt{1-\frac{b^{\dagger}b}{2j}b}\right)
\label{eq.HP_DH}
\end{eqnarray}
In the super-radiant phase to capture the macroscopic occupations of both the field and the atomic ensembles we have to displace the bosonic modes in Holstein-Primakoff, in either of the following ways.
\begin{eqnarray}
a^{\dagger}\rightarrow c^{\dagger} + \sqrt{\alpha};b^{\dagger}\rightarrow d^{\dagger} - \sqrt{\beta}\\
a^{\dagger}\rightarrow c^{\dagger} - \sqrt{\alpha};b^{\dagger}\rightarrow d^{\dagger} + \sqrt{\beta}
\label{eq.displacement}
\end{eqnarray}
In the DH obtained we retain only the terms linear in $j$. Both the choices of the bosonic displacements give identical Hamiltonians. Hence, every state is doubly degenerate in the super-radiant phase.\\
Diagonalising the Hamiltonian in the uncoupled $(q_{1},q_{2})$ basis we obtain the ground states as:
\begin{equation}
\Psi_{G}(q_{1},q_{2}) = G_{-}(q_{1})G_{+}(q_{2})
\label{eq.ground_state}
\end{equation}
 In this scheme the ground states in both the phases have a Gaussian profile ($G_{\pm}$, with different $(q_{1},q_{2})$ in both the phases), given in the artificial $(x,y)$ basis by:
\begin{eqnarray}
g(x,y)&=&\left(\frac{\epsilon_{+}\epsilon_{-}}{\pi^{2}}\right)^{1/4}Exp\left[\frac{-<\bf{R},A\bf{R}>}{2}\right]\\\\ \nonumber
A&=&U^{-1}MU\\
M&=&diag\left[\epsilon_{-},\epsilon_{+}\right]
\label{eq.Fid_red}
\end{eqnarray} 
$A$ is the rotation matrix parametrized with the angle $\gamma$ which is needed to transfer the basis from $(q_{1},q_{2})$ to $R = (x,y)$ and $U$ is an orthogonal matrix. $\epsilon_{\pm}$ are the atomic and the photonic excitations of the DH.
The ground state fidelity is given by:\cite{zanardi06,tapo}
\begin{equation}
\braket{g|g^{'}}=2\frac{\left[\det A \det A^{'}\right]^{1/4}}{\left[\det\left(A+A^{'}\right)\right]^{1/2}}
\label{eq.analytical_Fid}
\end{equation} 
which on simple determinant manipulation gives:
\begin{equation}
\braket{g|g^{'}}=2\frac{\left[\det M \det M^{'}\right]^{1/4}}{\left[\det\left(M+M^{'}\right)\right]^{1/2}}
\label{eq.analytical_Fid_M}
\end{equation}
This yields a fidelity expression for the normal phase as a function of the parameter $\lambda$ . In the plot of ground state fidelity vs the $\lambda$ in the thermodynamic limit we have colour coded the two phases differently as they arise from two different representations of the same Hamiltonian, on either side of the QPT. 

\begin{figure}[h]
{\includegraphics[width=8cm,height=4cm]{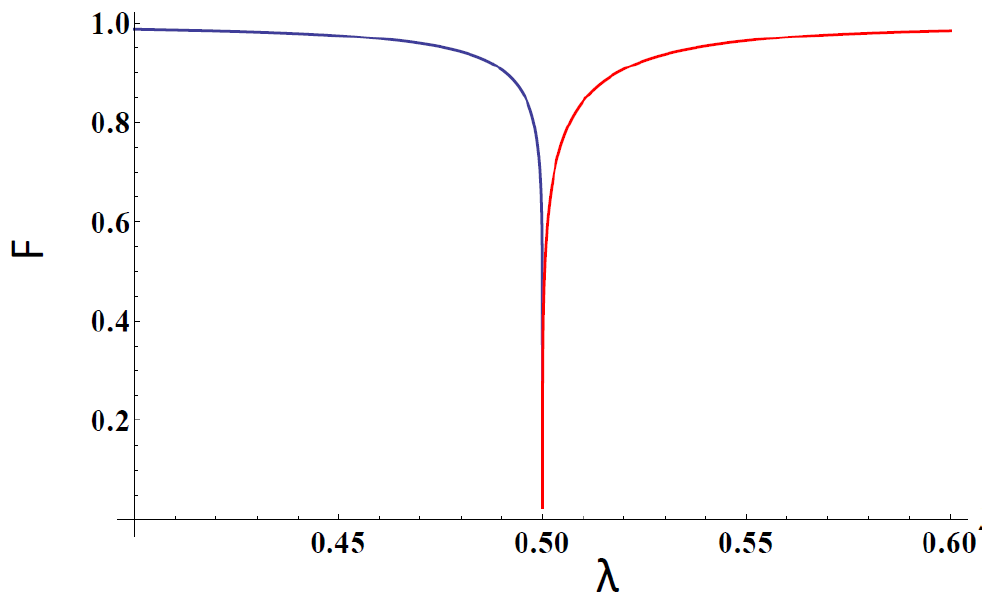}}
\caption{(colour online) The plot for the exact fidelity expression in the thermodynamic limit; one finds a sharp dip at the quantum critical point $\la_c=1/2$.}
\label{fig_Infin_fid}
\end{figure}

\subsection{Finite $j$}
To obtain the numerical value of fidelity we set the parameter $\delta =0.1$ and diagonalise the DH on either side of the critical point. We numerically obtain the ground states $\ket{\psi(\lambda)_{g}}$ and $\ket{\psi(\lambda + \delta)_{g}}$ to calculate the fidelity as defined earlier. 
 The fidelity when plotted against  $\lambda$ shows a dip near the thermodynamic QCP while  the slight difference is due to the finite size of the system.
 
 An immensely interesting behaviour of the fidelity occurs in the super-radiant phase at $\lambda >\lambda_{c}$. We see a significant number of oscillations in the fidelity which drops from a value less than unity to near zero. It rises and falls aperiodically till a value of $n_{c}$ when one can longer consider it as an appropriate bath. Then, the fidelity rises to one but the aperiodic oscillations persist. On increasing the value of $n_{c}$ (i.e., the size of the bath), we observe that the fidelity remains less than one up to even larger value of $\lambda$  though the oscillation persists. Ideally, an infinite bath size would see the fidelity never rise to one at any finite value of $\lambda$. In the $j\rightarrow\infty$ limit, the DH is integrable in both its phases and we recall the absence of aperiodic oscillations in both the phases of the plot at all values of $\lambda$ as shown in Fig.~(\ref{fig_Infin_fid}).
 
\begin{figure}[h]
\includegraphics[scale=.35]{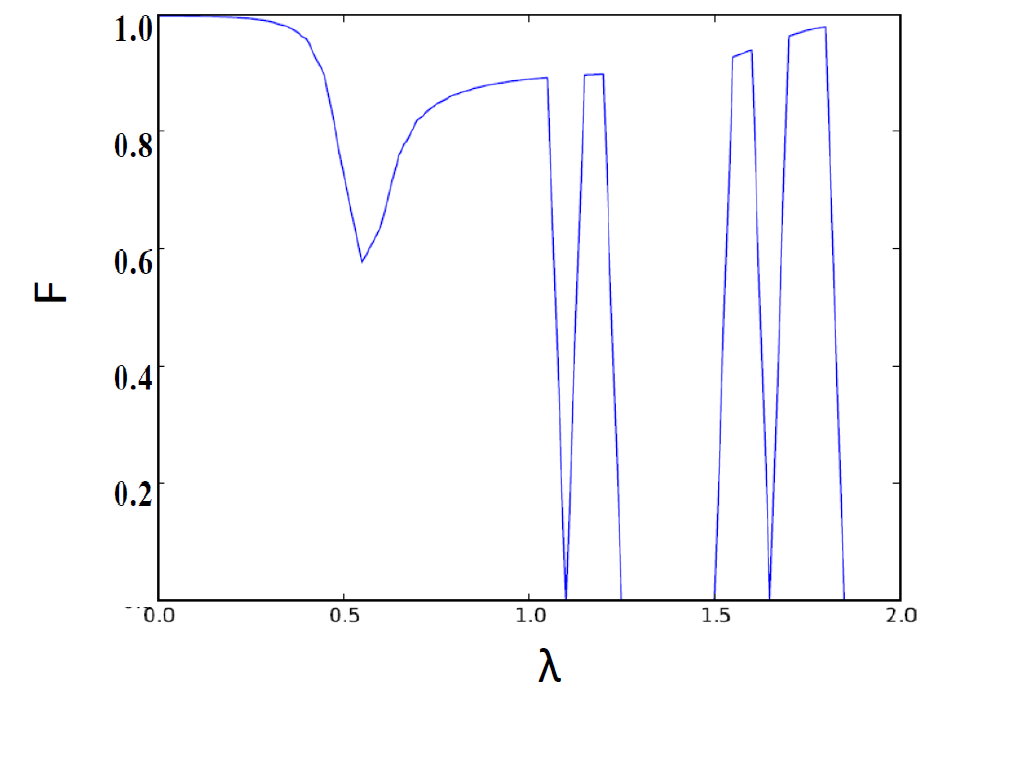}
\begin{picture}(0,0)
\put(-76,60){\includegraphics[height=2.6cm,width=3.0cm]{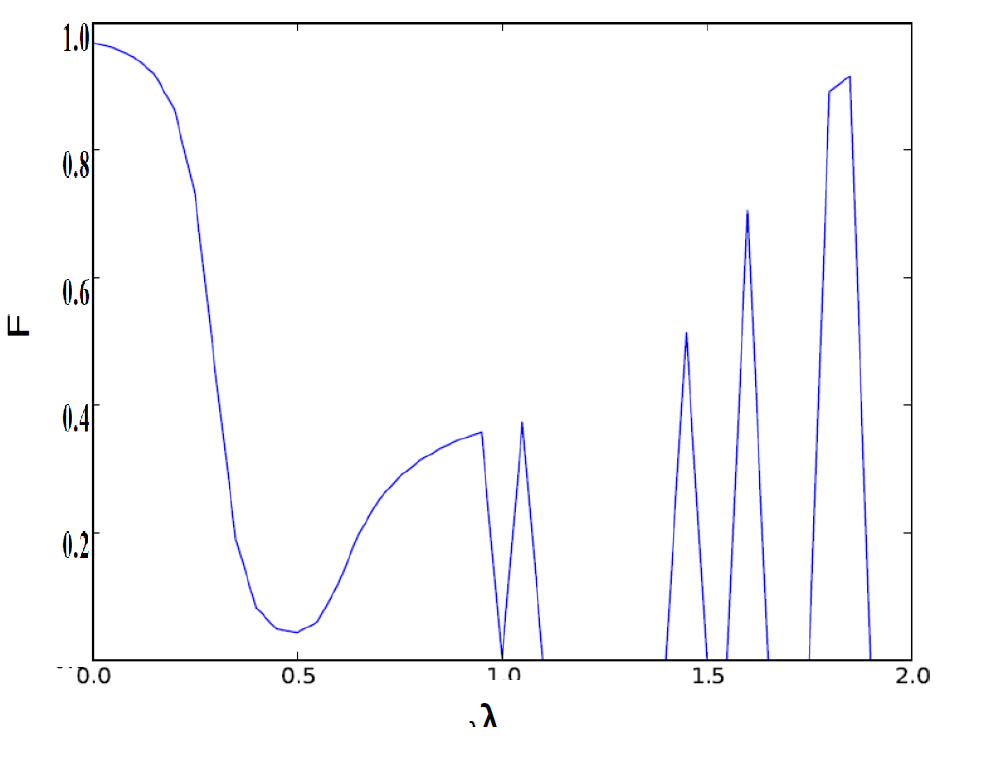}}
\end{picture}
\caption{(colour online) Fidelity for $\delta=0.1$ and $\delta =0.3$ (inset) for $n_{c} = 40$. As evident the ground state fidelity oscillates wildly as the system crosses $\lambda_{c}$ into the super-radiant phase. Inset: The plot in the case of a large deviation $\delta = 0.3$ clearly shows a lower recovery for the ground state fidelity}
\label{fig_fidelity_fin_1}
\end{figure}

\begin{figure}[h]
\includegraphics[scale=.34]{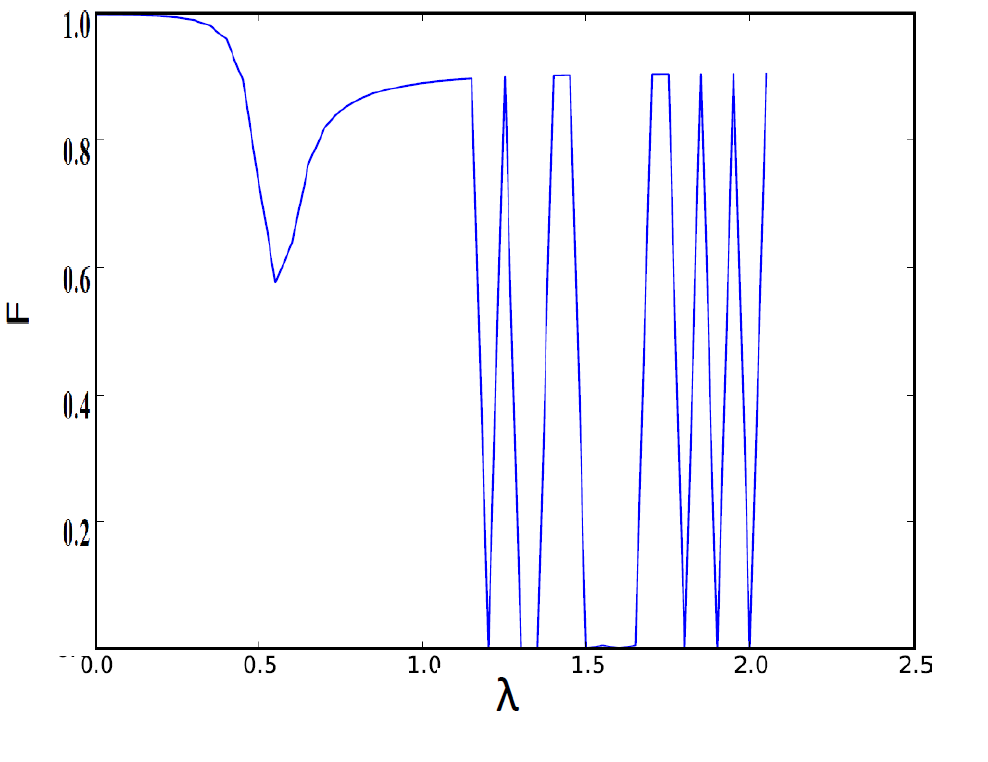}
\begin{picture}(0,0)
\put(-90,43){\includegraphics[height=3.2cm,width=2.8cm]{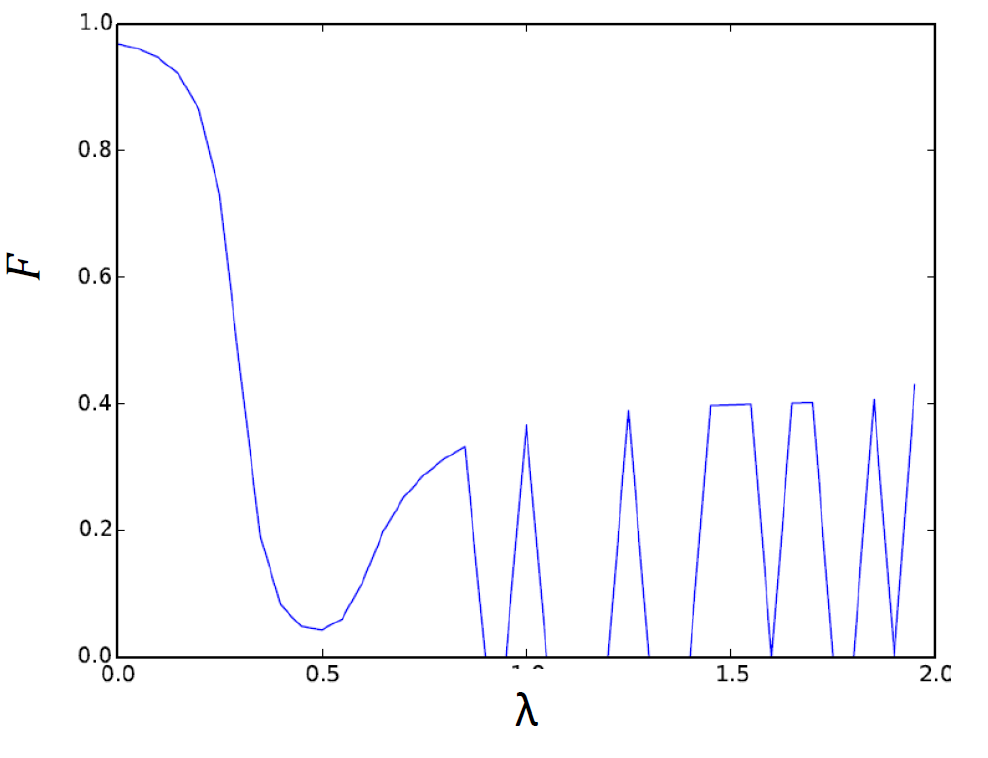}}
\end{picture}
\caption{(colour online) Fidelity for $\delta=0.1$ and $\delta =0.3$ (inset) for $n_{c} = 70$. The plot in the case of a larger bath size shows that the fidelity remains less than one for larger range of $\lambda$ without any change in the oscillatory behaviour in the super-radiant phase.}
\label{fig_fidelity_fin_2}
\end{figure}

Remarkably, the presence of chaos in the super-radiant phase, as indicated by the level crossing arguments and their statistics \ct{brandes}  manifests itself in the fidelity as aperiodic oscillations.
Even for a small change in the parameter $\delta$ in the Hamiltonian, we find that the ground states are  widely separated for some specific values of $\lambda (>\lambda_c)$ resulting in a nearly vanishing fidelity. For other values of $\lambda(>\lambda_c)$ also,  the overlap is small and decreases further with increasing $\delta$ seen in the insets of Figs.~[\ref{fig_fidelity_fin_1}] and [\ref{fig_fidelity_fin_2}]. Thus, unlike the normal phase where the fidelity remains very close to unity throughout with a dip at the critical point, one finds a remarkably
different behavior in the super-radiant phase.

\section{ Loschmidt Echo for finite $j$}
\label{sec_GroundstateLE}
\begin{figure}[h]
\includegraphics[scale=.3]{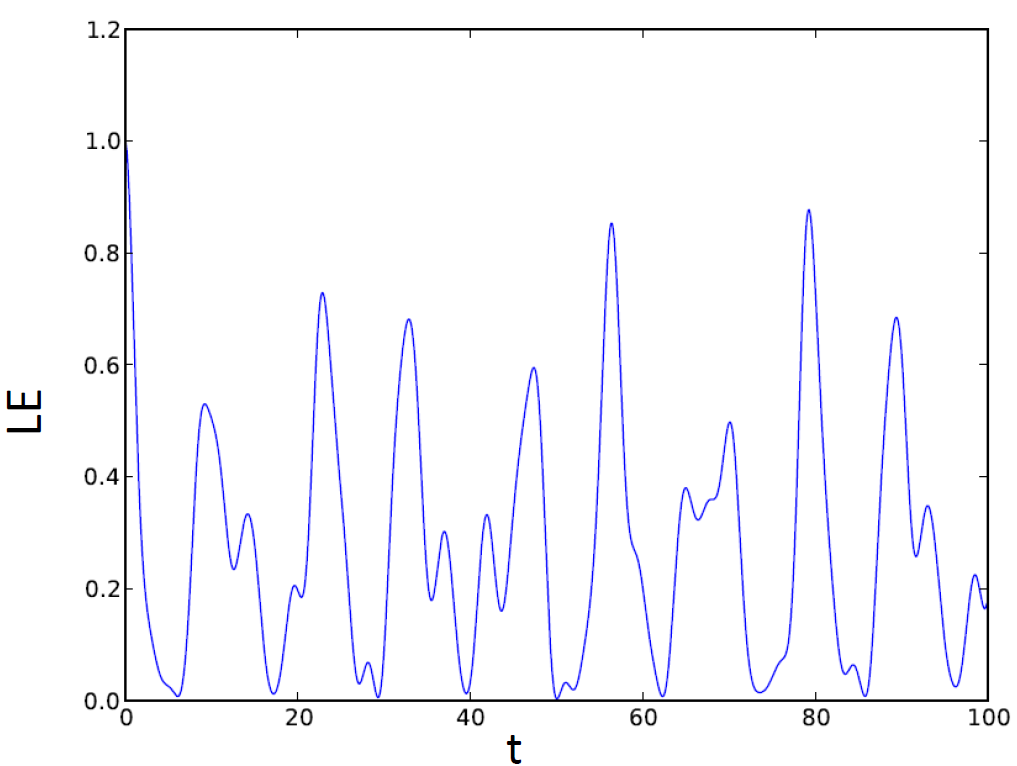}
\begin{picture}(0,0)
\put(-102,95){\includegraphics[height=2.2cm]{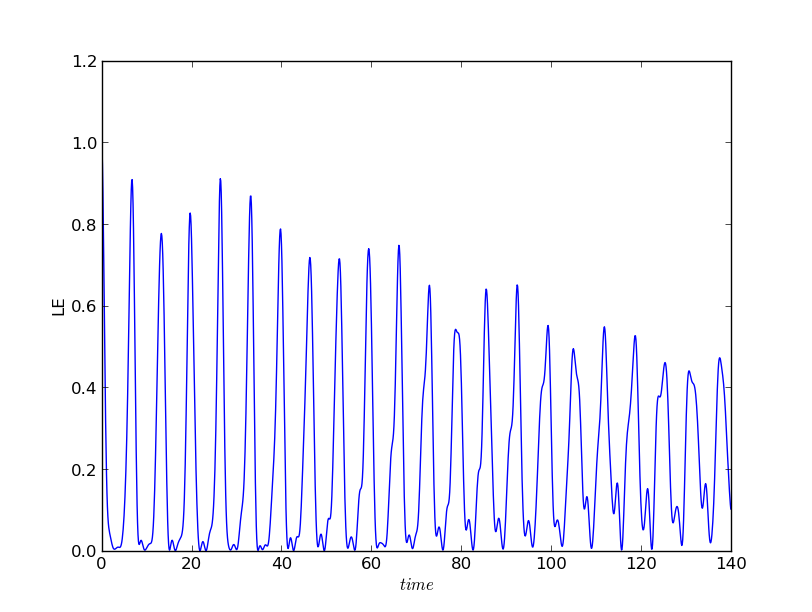}}
\end{picture}
\begin{picture}(0,0)
\put(-202,95){\includegraphics[height=2.2cm]{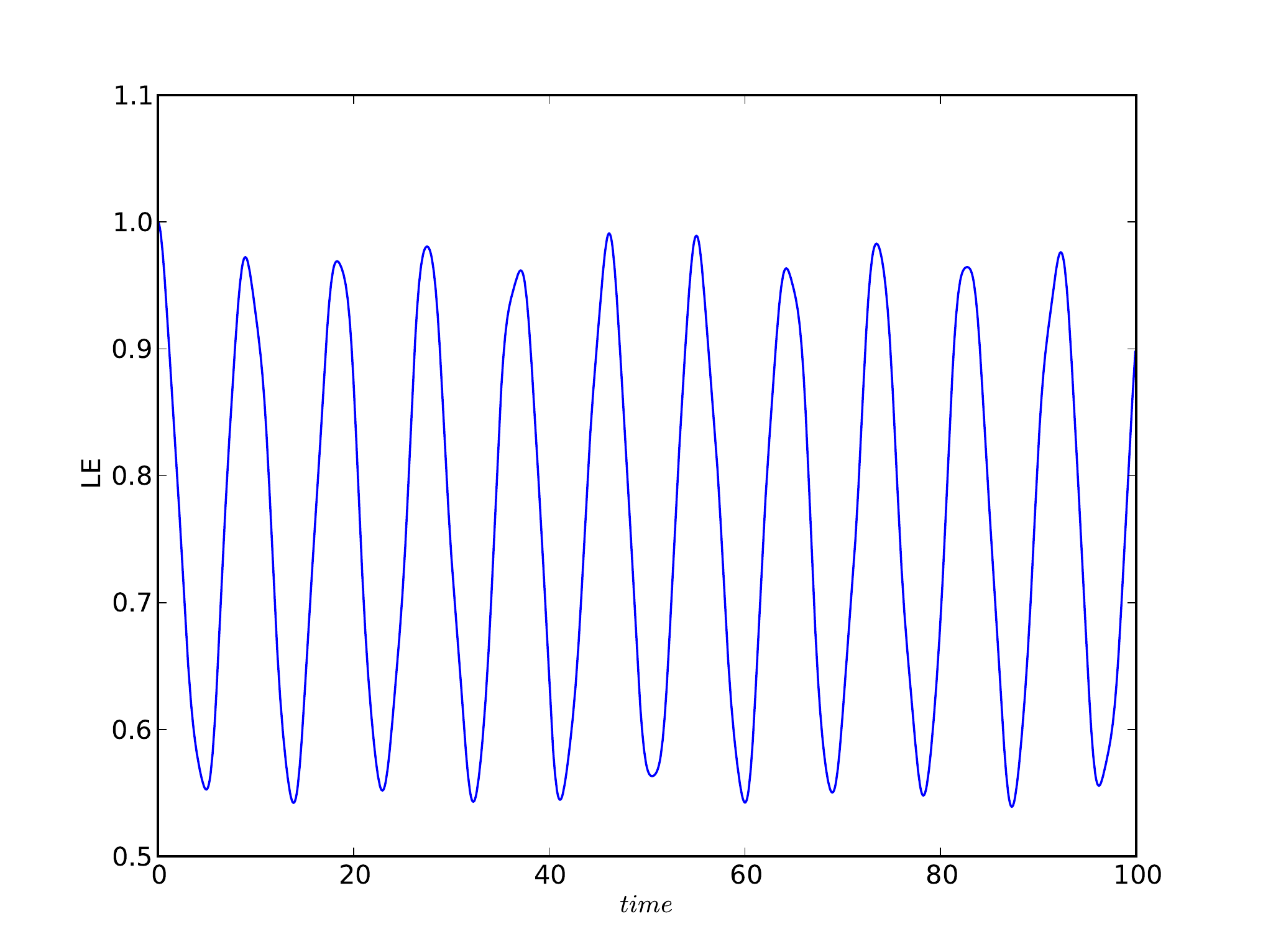}}
\end{picture}
\caption{(colour online) The Loschmidt Echo for finite $j$ and with $\delta = 0.3$ at a set time in the normal phase (top-left), at the QCP ($\lambda = \lambda_{c}$) (background) and in the super-radiant phase (top-right).The LE in the normal phase shows sustained oscillations (periodic) as only a few states are involved, as we move into the super-radiant phase the nature of the LE becomes aperiodic with many states contributing to the LE.}
\label{fig_LE_fin}
\end{figure}

The modulus of the overlap between the two ground states where one is evolved with $H(\lambda)$ and the other with a  shifted parameter $\lambda+\delta$ is known as the Loschmidt Echo (LE) given by the expression
\begin{equation}
L(t)=|\bra{\psi(\lambda)}e^{iH(\lambda)t}e^{-
iH(\lambda+\delta)t}\ket{\psi(\lambda)}|^{2}
\label{eq.LE}
\end{equation}
We study the time evolution of the  LE  in the  normal phase and the super-radiant phase as well as at the QCP for appropriate values of the parameter $\la$. We list the observations below:

In the normal phase Fig.~\ref{fig_LE_fin}(top-left), we find  that the amplitude of the LE varies from a value of 1.0 to 0.55 and the peaks in the envelop have nearly the same amplitude.
Near the QCP Fig.\ref{fig_LE_fin} (background), the ground states at $\lambda_{c}$ and at $\lambda_{c} +\delta$ are widely separated, hence, we see that the LE dips from 1 to 0 and theres no apparent periodicity marking the QCP at around $\lambda=0.5$.
In the super-radiant phase Fig.\ref{fig_LE_fin} (top-right), there is an overall decay in the amplitude of the LE with time. The amplitude of the envelop revives after a long time.

\begin{figure}[h]
\includegraphics[height = 6cm]{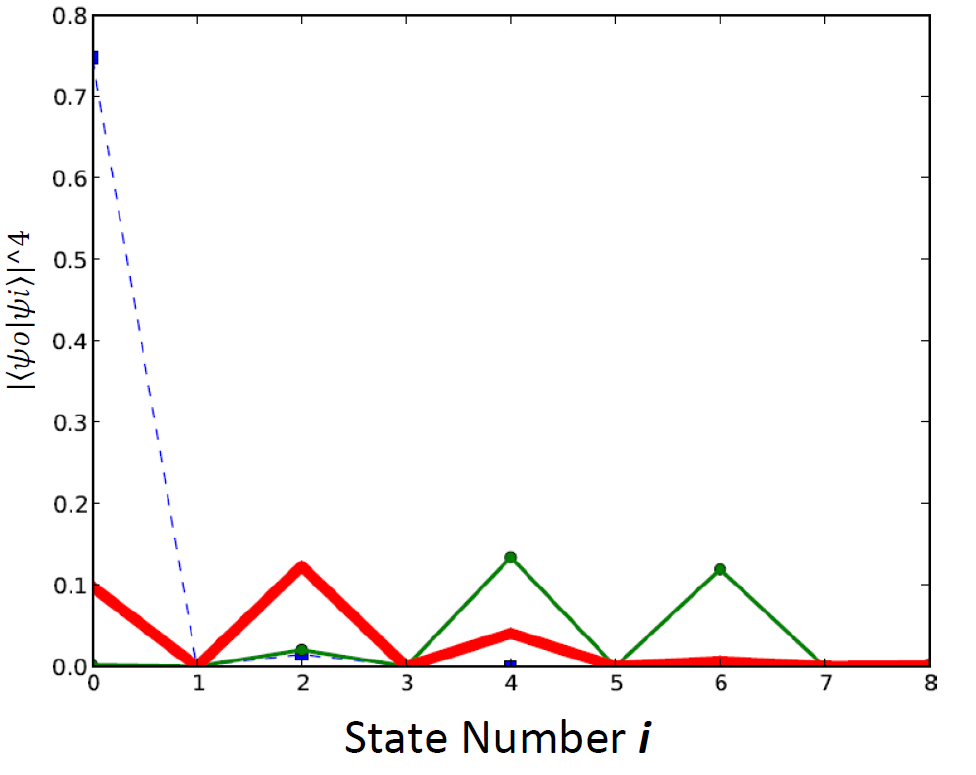}
\caption{(colour online) This plot shows the mod square of the overlap between $\bra{\psi_{o}(\lambda)}$ and $\ket{\psi_{i}(\lambda +\delta)}$ for $i\in[0,n]$ corresponding to the LE plots: blue (dotted) for the normal phase, green (thin-line) at the QCP and red (thick-line) for the super-radiant phase. One can clearly observe the delocalisation taking place in the state space with increase in $\lambda$.}
\label{fig_overlap}
\end{figure}

We see from Fig.\ref{fig_overlap}, which is a plot of the overlap between the ground state at $\lambda$ and all states at $\lambda_{c} +\delta$ against the total number of states, that in the normal phase the overlap between the ground state and the states of the Hamiltonian with a shifted value of $\lambda$ is limited to one or two excited states. Hence there is no decay of the LE with time, the system aperiodically oscillates with the superposition of two or three frequencies associated with the energy differences of the non zero overlaps.
In the super radiant phase, in contrary, we see a delocalization of the wave function with parameter $\lambda + \delta$. As evident from a greater number of states of $\ket{\psi_{i}(\lambda + \delta)}$) contributing to the overlap with the ground state with smaller amplitudes, than in the normal phase. As a larger number of overlaps are involved the phases interfere destructively leading to a decay of the LE with time.
Finally at the crossover point $\lambda_{c}$ we see a mixture of both the above mentioned behaviors: chaotic and non-chaotic regimes of $\lambda$ get involved in the LE and so we get an aperiodic pattern.

\section{The Time-Average of LE}
\label{sec_TimeavgLE}
Generally, the LE serves as a good indicator of QCP, but  to understand the transition to chaos in the super-radiant phase of the DH, one should explore the time average of the LE as argued by Peres\cite{peres84}. It has been suggested that if a quantum system has a chaotic classical analogue then the time average of the overlap between two states  nearly
vanishes in the chaotic phase while it remains close to unity in the regular phase. We employ the same technique in the present context using the two ground states evolved with two slightly different Hamiltonians. It is easy to show that:
\begin{equation}
\braket{L}= \lim_{T\rightarrow\infty}\frac{\int_{0}^{T}L(t)dt}{\int_{0}^{T}dt}=\sum_{i}|\braket{\psi_{o}|\psi_{i}}|^4,
\label{eq.LE_avg}
\end{equation}
where $\psi_{0}$ is the ground state of $H(\lambda)$ and $\psi_{i}$ is the $i^{th}$ excited state of the  Hamiltonian with
the modified value of $\lambda$. We emphasize that though there is an apparent similarity with the expression for fidelity, there is also a subtle difference, this expression incorporates information about all the excited states
of the Hamiltonian $H(\lambda+\delta)$. Thus the time average LE  is expected to  capture the entire delocalisation scheme unlike the fidelity.

\begin{figure}[h]
\includegraphics[height=5.8cm,width = 7.8cm]{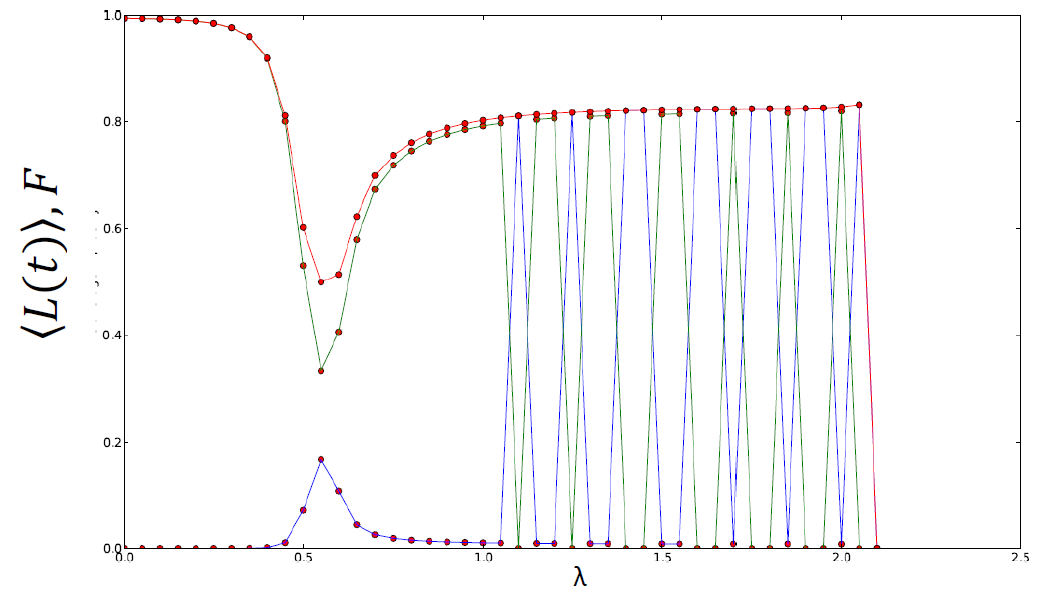}
\caption{(colour online) A combined plot of the time average LE (red), ground state fidelity (green), complementary higher state fidelity sum (blue) versus $\lambda$. It can be clearly seen that the time average of the LE contains within it information about the ground state fidelity as expected; making the picture of delocalization of $\ket{\psi}$ in state space clear as we move into the super-radiant phase.}
\label{fig_plot_con}
\end{figure}

A simple mathematical expression connects the time averged LE and the ground state fidelity:
\begin{equation}
\label{eq.connection}
\braket{L} = F^{2} + \sum_{i\neq0}\braket{\psi_{i}|\psi_{0}}^{4}.
\end{equation}
Figure \ref{fig_plot_con} clearly shows that the first dip of the ground state fidelity (green)  in the chaotic phase occurs at the a value of $\lambda$  where the LE average (red) just starts to flatten out. This implies that the terms with $i\neq0$ 
in Eq.~(\ref{eq.connection}) oscillates complementary to that of the square of the fidelity, clearly showing a clear connection between the fidelity and the time averaged LE. Thus we can conclude  that the LE average already
incorporated  the effect of ground state fidelity while providing a  clearer picture of the delocalisation in state space.\\

The time average of LE distinctly separates the two phases of the DH. As argued in \ct{peres84}, the occurrence of chaos in the super-radiant phase is indicated by the time average dipping to a value much less than one, whereas in the normal phase, the time average remains close to unity indicating regularity. This is because in the normal phase, on slight change of the perturbing parameter $\delta$, only a few excited states near the ground state of the unperturbed Hamiltonian are occupied where the   overlap with initial ground state becomes significant; this indicates that the wave function remains localized, which should indeed be the case for regular behavior.

In the chaotic phase, the delocalisation of the wavefunction can be understood as a signature of chaos, since, on slight change of $\delta$ here, a large number of the excited states get occupied and resultantly, the overlap with the unperturbed ground state becomes small. In classical picture chaos is understood as the exponential separation of two trajectories with very similar initial conditions. This effect is manifested  in the quantum analogue, by distribution over the state space of two states with slightly different Hamiltonian parameters. As the time average, of LE contains this distribution through the sum over all states, we see a significant drop in the value in the super-radiant phase clearly pointing to delocalisation and hence  a signature of chaos.

\begin{figure}[h]
\includegraphics[scale=.3]{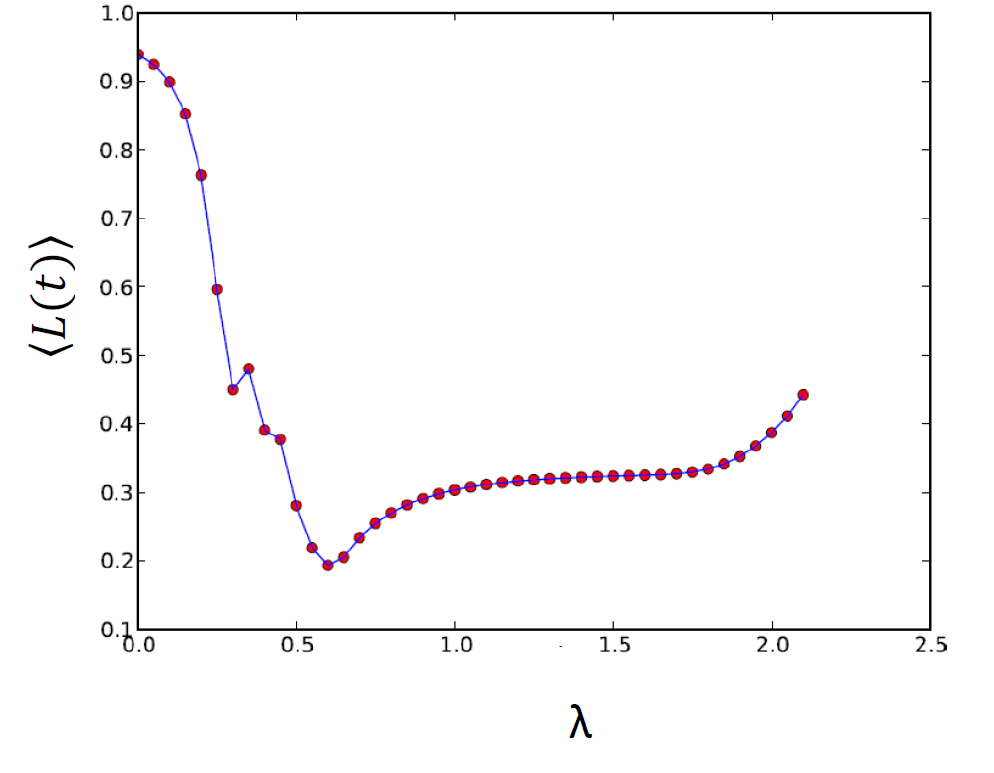}
\begin{picture}(0,0)
\put(-122,70){\includegraphics[height=2.8cm,width = 3.4cm]{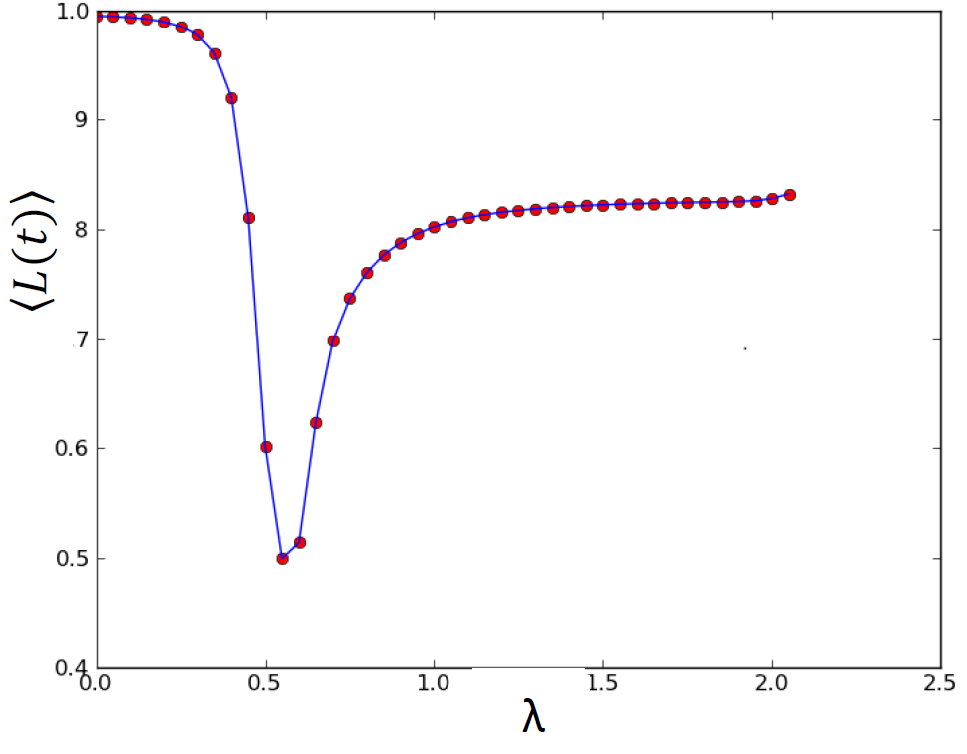}}
\end{picture}
\caption{(colour online) Time averaged Loschmidt Echo for $\delta =0.3$ (background) and $\delta = 0.1$ (inset). In both the cases the time average of the LE dips significantly from 1 as $\lambda$ goes into the chaotic super-radiant phase. Inset: Like in the case of ground state fidelity (Fig.\ref{fig_fidelity_fin_1}) as $\delta$ is increased the recovery of the time average of the LE to 1 never really occurs.}
\label{fig_LE_avg_fin}
\end{figure}

\section{Conclusion}
\label{conclusion}
We have used Loschmidt Echo and its average to study chaos in the Dicke Hamiltonian with finite $j$ employing the resonance condition throughout. We have also observed that the ground state fidelity shows a similar behaviour through the random oscillations in the super-radiant phase.

In our case we concentrate on the delocalisation produced in the state space by the onset of chaos as compared to the distribution of eigen energies as studied in earlier works. We conclude that the delocalisation of the wavefunction manifests itself in the time average of LE and the fidelity giving clear signs of the presence of chaos in both the quantum and the semi-classical Hamiltonian.

\end{document}